\documentclass[letterpaper]{JHEP3}
\usepackage{epsfig,enumerate}
\usepackage{subfigure}
\usepackage{amsmath, amssymb, graphics}

\newcommand{\mathsym}[1]{{}}

\newcommand{\eref}[1]{(\ref{#1})}

\renewcommand\({\left(}
\renewcommand\){\right)}
\renewcommand\[{\left[}
\renewcommand\]{\right]}

\newcommand{\dd}{{\rm d}}
\newcommand{\e}{{\rm e}}

\newcommand\eps{\epsilon}
\newcommand\mpl{m_{\rm p}}

\def\ba{\begin{eqnarray}}
\def\ea{\end{eqnarray}}
\def\be{\begin{equation}}
\def\ee{\end{equation}}

\def\O{\mathcal{O}}

\def\R{\mathcal{R}}
\def\M{\mathcal{M}}

\def\nn{\nonumber}
\def\({\left(}
\def\){\right)}

\def\eref#1{(\ref{#1})}

\newcommand{\roughly}[1]{\mathrel{\raise.3ex\hbox{$#1$\kern-0.85em
\lower1ex\hbox{$\sim$}}}}

\title{A no-go for no-go theorems prohibiting cosmic acceleration in extra
  dimensional models}

\author{Rik Koster$^{\,a,b}$ and Marieke Postma$^a$ \\$^{\,a}$ Nikhef, Science Park 105, 1098
XG Amsterdam, The Netherlands. \\
$^b$ Department of Physics and Astronomy, VU University, De Boelelaan
1081,\\ 
$^{\,}$ 1081 HV Amsterdam}

\date{}

\abstract{A four-dimensional effective theory that arises as the
  low-energy limit of some extra-dimensional model is constrained by
  the higher dimensional Einstein equations.  Steinhardt \& Wesley use
  this to show that accelerated expansion in our four large dimensions
  can only be transient in a large class of Kaluza-Klein models that
  satisfy the (higher dimensional) null energy condition \cite{SW}.
  We point out that these no-go theorems are based on a rather ad-hoc
  assumption on the metric, without which no strong statements can be
  made.}

\preprint{Nikhef 2011-025}

\begin{document}

\section{Introduction}

There is compelling evidence that both the early universe as well as
our present-day universe went or is going through a phase of
accelerated expansion, referred to as inflation and dark energy
respectively. The evolution of the universe is sourced by the
energy-density in it; acceleration requires the energy-density to be
dominated by vacuum energy. This cannot be achieved by ordinary
matter, however, an (effective) scalar field whose kinetic energy is
negligible will work.  In the context of extra-dimensional models such
a four-dimensional (4d) scalar can either arise from higher
dimensional matter, or as components of the extra-dimensional metric
(called moduli fields).  The higher dimensional fields should satisfy
the higher dimensional Einstein equations.  This puts constraints on
the matter properties of the low-energy effective four-dimensional
theory.

In Ref. \cite{SW} (and also in the earlier Refs. \cite{oxidised,
  wesley}) Steinhardt and Wesley use these constraints to derive
stringent no-go theorems on four-dimensional accelerated expansion in
models with extra dimensions.  It was for example shown that if the
higher dimensional theory satisfies the null energy condition (NEC),
accelerated expansion can only last for a couple of e-folds. Models
that do violate the NEC can yield accelerated expansion, but the time
and spatial dependence of the NEC violating elements is heavily
constrained. To derive these theorems, it is assumed that both the
higher dimensional and the 4d compactified theory are described by
general relativity (up to small corrections).  The metric is taken
block-diagonal, with the extra-dimensional part Ricci flat or
conformal Ricci flat.  This form is motivated by many models that
exist in the literature, such as the original Kaluza-Klein model,
Randall-Sundrum models \cite{rs1,rs2}, and string compactifications on
(warped) Calabi-Yau manifolds \cite{GKP, KKLT, LVC, klebanov}.

The no-go theorems derived in Ref.~\cite{SW} seem to forbid dark
energy and inflation\footnote{If during inflation the corrections to
  the Friedmann equation in the dimensionally reduced theory are
  large, the theorems may be avoided.} in many motivated
extra-dimensional models. However, every no-go theorem is as good as
the assumptions that go into it. In this paper we point out some of
these assumptions, which are non-trivial and key ingredients in the
derivation of the no-go theorems. None of these are mentioned
explicitly in Ref. \cite{SW}, although they can be found in some form
in the earlier works\cite{oxidised,wesley}.

First of all, the theorems derived in Ref.~\cite{SW,oxidised,wesley}
only apply to Kaluza-Klein (KK) type compactifications, and not to a set-up
with branes or, more generally, localized matter sources. There are
two instances where brane sources may violate the assumptions made.
First, the theorems are formulated using a partial integration, where
it is assumed the boundary term vanishes:
\be
\int_{\partial \M} \e^{(2+A)\Omega} (\vec\nabla \Omega). \dd
\vec\Sigma = 0.
\nn
\ee
Here $\Omega$ is the warp factor, $\vec \nabla$ the gradient
constructed from the extra-dimensional metric, $A$ a constant, and
$\vec\Sigma$ the oriented boundary of compactified space $\M$.  This
is automatic if $\M$ is closed and has no boundaries (for orbifold
compactification, if the covering space has no boundaries), as in KK
compactifications. However, if $\M$ is bounded by branes, boundary
terms as the above are generically non-zero, as they are sourced by
the brane localized energy-momentum, and the no-go theorems do not
apply \cite{nierop2}.  Second, it is assumed that the metric of our
universe $g^{\rm vis}_{\mu\nu} $ equals the metric $g^E_{\mu\nu}$ that
brings the dimensionally reduced 4d Einstein-Hilbert action in
canonical form. This is not automatic in brane world models.  Indeed,
if the standard model matter is localized on a brane instead of living
in the bulk the induced metric on the brane, which is the metric of
our universe, is
\be
g^{\rm vis}_{\mu\nu} = \e^{2\Omega(y_{\rm br},t)} g^E_{\mu\nu},
\nn
\ee
with the warp factor evaluated at the position of the brane $y_{\rm
  br}$. We can normalize this to $\Omega(y_{\rm br},0) =1$ by
rescaling the the 4d coordinates. Only if the warp factor is
time-independent the two metrics coincide at all times.  Otherwise
accelerated expansion in our universe is not the same as accelerated
expansion in the canonical metric.  The two metrics enter because
matter is localized to the brane, whereas gravity lives in the bulk.

Secondly, even for KK type compactifications the theorems are not as
general as suggested in Ref.~\cite{SW}, as an additional restriction
is put on the metric, which relates the warp factor to the volume
modulus of the extra dimensional space.  Although it is presented as a
necessary condition (which is enforced by hand) in earlier work
\cite{oxidised,wesley}, it should really be viewed as an additional
assumption.  This metric restriction is motivated in that an extra
constraint is needed to eliminate modes with a wrong-sign kinetic term
in the four-dimensional effective action.  However, the off-diagonal
Einstein equations, which are not taken into account in the derivation
of the no-go theorems, can provide exactly such constraints.  Another
argument given is that it is merely a gauge choice in the ``moduli
space approximation'' \cite{oxidised}.  But this approximation breaks
down in a time-dependent set-up, which is exactly what we are
interested in when looking for 4d accelerated expansion.  That the
metric restriction is indeed unnecessary is confirmed by an explicit
counter example: the solution of the five-dimensional Einstein
equations found in Ref.~\cite{gio} violates it.

No-go theorems forbidding a 4d de Sitter (dS) space already exist for
a long time \cite{gibbons,wit,maldacena}.  These initial works showed
that string theory or supergravity compactifications on a smooth
time-independent extra-dimensional manifold cannot give a 4d
cosmological constant if the original theory satisfies the strong
energy condition.  Since then much work has been done to generalize
and extend these results.  Negatively curved extra-dimensions are
discussed in \cite{douglas}.  Theorems forbidding dS solutions in
string theory constructions with orientifold planes, which break the
null energy conditions, have recently been formulated in
\cite{shiu,sethi}. In
\cite{scrucca1,scrucca2,covi1,zagermann,giddings} it was shown that
supergravity compactifications to dS often are unstable as they have a
tachyon in the spectrum.  Theorems restricting a period of 4d
inflation can be found in \cite{covi2,hertzberg}.  All these theorems
are formulated in a string theory or supergravity context, and
supersymmetry and/or knowledge of the matter content in string theory
(e.g. on the form of the K\"ahler potential) is used to derive them.
The approach of \cite{SW} is orthogonal in a way, as it does not rely
on supersymmetry, nor on the form of the matter in the theory except
that it satisfies the null energy condition.  Given that it is stated
so general, it may be no surprise that extra assumptions, such as the
metric assumption mentioned above, are needed to derive no-go
theorems.

This paper is organized as follows.  We start in the next section
reviewing the no-go theorems of \cite{SW}. To do so we use Einstein's
equations to relate the NEC in the higher-dimensional theory to the
metric moduli and four-dimensional scale factor. In section
\ref{s:assumptions} we take a step back, and list all the assumption
that went into the derivation.  We discuss in some detail how brane
world models generically evade one or more of the assumptions.  In
section \ref{s:restriction} we discuss the restriction put on the
metric in Ref. \cite{SW,oxidised,wesley}, that relates the warp factor
to the volume modulus of the extra dimensional space.  We argue that
it should be viewed as an assumption rather than a necessary
restriction.  We further show that without this assumption no general
no-go theorems that can be derived.  We end with some concluding
remarks.

\section{Einstein's equations and the null energy condition}
\label{s:equations}

In this section we review the no-go theorems of \cite{SW}. As a
preliminary we introduce the metric and define the 4d Einstein frame,
and define an averaging procedure.  We then use the Einstein equations
to relate the average NEC violation in the higher dimensional theory
to the 4d scale factor and moduli fields.  We use a slightly different
gauge than \cite{SW}, setting $\phi =0$ in their equations, to make
the derivations as transparant as possible.

\subsection{The metric}

We assume the metric to be of the block-diagonal form
\be
\dd s^2 = g^{(D)}_{MN} \dd X^M \dd X^N
=
\e^{2\Omega(t,y)} g^E_{\mu\nu}(t) \dd  x^\mu \dd x^\nu
+ g^{(d)}_{\rm mn}(t,y) \dd y^m \dd y^n,
\label{metric}
\ee
with coordinates $X^M = (x^\mu,y^m)$.  Here ($M =0,..,D-1$) runs over
all $D$ dimensions, whereas ($\mu =0,..,3$) labels our four large
space-time dimensions and ($m = 4,...,D-1$) the $d = D-4$ extra
dimensions. The four-dimensional Einstein frame metric $g^E_{\mu\nu}$
is of the Friedmann-Robertson-Walker (FRW) form
\be
g^E_{\mu\nu}(t) \dd  x^\mu \dd x^\nu = - n(t)^2 \dd t^2 +a^2(t) \dd \vec x^2,
\label{metric_FRW}
\ee
appropriate to describe (accelerated) expansion in a homogeneous and
isotropic universe.  We assume that the universe is flat, in
agreement with observations \cite{WMAP}.  In this metric $n(t)$ is the
lapse function and $a(t)$ the scale factor.  Accelerated expansion is
defined by (taking $a>0$)
\be
\frac{\ddot a}{a} = \dot H + H^2> 0
\label{dda}
\ee
with $H= \dot a/a$ the Hubble constant.  The extra-dimensional metric
is taken of the form
\be
g^{(d)}_{mn}(t,y)  = \e^{-2\bar \Omega(t,y)} \bar g^{(d)}_{mn}(t,y), 
\label{RF}
\ee
with the scalar curvature constructed from the barred metric vanishing
$\R[\bar g_{mn}^{(d)}] = 0$.  The metric is Ricci flat (RF) if $\bar
\Omega$ is constant (can be set to zero by a redefinition of $\bar
g_{mn}^{(d)}$), and conformal Ricci flat (CRF) if $\bar \Omega =
\Omega$ equals the warp factor.  The Ricci scalar of the
extra-dimensional space is $\R[g^{(d)}_{mn}] = 0$ for a RF metric, and
\be
\R[g^{(d)}_{mn}] = 
2(d-1) \nabla^2 \Omega +(d-1)(d-2) (\nabla_m \Omega)^2, 
\label{R}
\ee
for CRF. Here $\nabla^2 = \nabla_m \nabla^m$, with $\nabla_m$ the
covariant derivative constructed from $g_{mn}^{(d)}$.

A generic coordinate transformation $X^M \to X^M({X'}^N)$ will alter
the form of the metric, in particular it will generate off-diagonal
terms $g_{\mu m} \dd x^\mu \dd y^m$.  Hence the block-diagonal form of
the metric \eref{metric} fixes most of the gauge. Residual invariance
is left of the form $t \to t(t')$, $\vec x \to \vec x(\vec x')$, $y^m
\to y^m({y'}^n)$.  The first transformation can be used to set
$n(t)=1$, which we will do in the following.  The second can be used
to fix the scale factor at a particular time, it is customary to set
the scale factor today to unity $a(t_{\rm now}) = 1$. We discuss the
third below.

We decompose the (time-derivative) of the extra dimensional metric
according to \cite{SW}
\be
\frac12 \partial_t g^{(d)}_{mn}(t,y) = \frac1d \xi(t,y)\, g^{(d)}_{mn}(t,y) 
+ \sigma_{mn}(t,y),
\label{dt_g}
\ee
with ${g^{(d)}}^{mn}\sigma_{mn} =0$ traceless.  The trace factor $\xi$
parametrizes the change in volume: $\partial_t \ln \sqrt{g^{(d)}} =
\xi$.  Under a time-independent coordinate transformation $y^m \to
y^m(y^n)$, which keeps the metric in block-diagonal form, $\xi$ is
invariant whereas $\sigma_{mn}$ transforms as a tensor.  The
non-invariance of $\sigma_{mn}$ will not be important for the
derivation of the no-go theorems, as this quantity only appears in a
covariant contraction in the equations.  However, if we want to solve
the constraint equations arising from the off-diagonal components of
Einstein's equations, we do have to worry about gauge invariance.

\subsection{A-averaging and the Einstein frame}

It will be useful to define an A-averaging procedure via
\be
Q_A(t) \equiv \langle Q(t,y) \rangle_A = 
\frac{\int Q \e^{A \Omega} \sqrt{g^{(d)}} \dd^d y}
{\int \e^{A \Omega} \sqrt{g^{(d)}} \dd^d y},
\label{average}
\ee
where the weight of the average is set by $A$.  Using this average any
quantity can be split in a y-independent zero-mode piece and the
fluctuations around it
\be
Q(t,y) = Q_A(t) + Q_{A\perp}(t,y)
\label{split}
\ee
with $\langle Q_{A\perp}\rangle_A =0$.
Splits using different values of $A$ only differ by a time-dependent
function: $Q_A(t) = Q_B(t) + f_{AB}(t)$.  The averaging procedure is
time-dependent, and in general the time-derivative of an average
differs from the average of a time-derivative:
\be
\partial_t \langle Q \rangle_A = \langle \dot Q \rangle_A
+ \langle Q (A \dot \Omega+\xi)_{A\perp}  \rangle_A.
\label{dQ}
\ee

The Einstein frame is found by integrating the action over the extra
dimensions
\be
S =\frac{1}{2\kappa^2_{_D}} \int \dd^{D} X \sqrt{-g^{(D)}} \R[g_{MN}^{(D)}]=
\frac{1}{2\kappa^2_{_D}} \int \dd^4x \sqrt{-g^E} 
 \(\int \e^{2\Omega} \sqrt{g^{(d)}} \dd^d y \) \R  [g^E_{\mu\nu}]
+...
\ee
with $\kappa^{-2}_D = M_D^{D-2}$ with $M_D$ the $D$-dimensional Planck
scale.  For $g_{\mu\nu}^E$ to be the four-dimensional Einstein metric
we need
\be
\frac{1}{\kappa^2_{_D}} \(\int \e^{2\Omega} \sqrt{g^{(d)}} \dd^d y \) = 
\frac{1}{\kappa^2_{4}},
\label{planckmass}
\ee
with $\kappa^{-2}_4 = \mpl^2$.  By definition, the Planck mass is
constant in the Einstein frame. Taking the time-derivative of the
above equation implies
\be
\langle 2\dot \Omega + \xi \rangle_2 = 0.
\label{einstein}
\ee
Note that the above relation is only valid for what we will call the
``canonical average'', with $A = 2$. Demanding $g^E_{\mu\nu}$ to be
the Einstein frame metric completely fixes $\Omega$ via
\eref{planckmass}.  There is no freedom left to rescale $\Omega \to
\Omega + c(t)$ and $g^E_{\mu\nu} \to \e^{-2c(t)}g^E_{\mu\nu}$.

\subsection{Einstein's equations}

Both the higher and four-dimensional energy and momentum are defined
in their respective Einstein frames $g_{MN}^{(D)}$ and $g^E_{\mu \nu}$.  
The definition of the D-dimensional energy density and pressure are
\be
\kappa^2_{_D} \rho = - g^{00}G_{00}, \quad\kappa^2_{_D} p_3
= \frac13 g^{ij} G_{ij},
\quad\kappa^2_{_D} p_d = \frac1d g^{mn}G_{mn}
\label{Energy}
\ee
with as before $D = 4+d$ the total number of dimensions.  A possible
higher-dimensional cosmological constant is absorbed in the energy and
pressure. The extra dimensions do not need to be homogeneous or
isotropic, and thus the extra dimensional part of the energy-momentum
tensor may have off-diagonal entries. Only the trace is used for the
theorems, but all components are needed to fully solve Einstein's
equations. The off-diagonal equations, which provide constraints, are
not used in the derivation of the theorems of Ref.~\cite{SW}.

We can dimensionally reduce the theory to four dimensions.  If the
system is to describe our universe today, this should yield general
relativity plus  corrections that encode its extra-dimensional
origin.  Hence, we get the usual Friedmann equations relating the
scale factor $a(t)$ to the effective energy and pressure plus 
corrections that are small in the low energy limit:
\be
\(\frac{\dot a}{a}\)^2 = \frac{\kappa^2_4}{3}\rho^E +...,
\qquad
 \(\frac{\ddot a}{a}\) =  - \frac{\kappa^2_4}{6} \rho^E(1 +3w^E)+...,
\label{FRW}
\ee
where we introduced the equation of state parameter $p^E = w^E
\rho^E$. The energy and pressure is the {\it total} four-dimensional
energy and pressure, which includes the dimensional reduction of any
matter added in the higher dimensional theory (possibly localized on
e.g. a brane), as well as the energy and pressure of the metric moduli
fields.   

At low energies, the scales being probed are large compared to the
typical size of the extra dimensions, and the corrections to the FRW
equations (denoted by the ellipses) are small. We know from
observations that during nucleosynthesis, when the temperature in the
universe was about an MeV, the corrections to the Friedmann equations
should be small. Extrapolating, they are completely negligible
today. It then follows that accelerated expansion requires dark energy
with an equation of state parameter $w_E < −1/3$. However, this is not
necessary the case for inflation, as this period of accelerated
expansion takes place well before nucleosynthesis where the
corrections to the Friedmann equations may be large. For example, in
RS models there is an additional $\rho^2$ term on the right-hand-side
of the first Friedmann equation which may play a r\^ole during
inflation \cite{FRWbrane1,FRWbrane2,FRWbrane3,FRWbrane4}. In this
regime statements on the scale factor or Hubble parameter (as in
(2.3)) cannot be straightforwardly translated in a statement on the
equation of state parameter. For this reason we will state our
theorems in terms of the scale factor, rather than the equation of
state parameter.

The D-dimensional null energy condition (NEC) states that $T_{MN} n^M
n^N \geq 0$ for any null vector $n^M$. The NEC is satisfied by all
unitary two-derivative quantum field theories. Although NEC violating
theories in general suffer from difficulties with unitarity,
superluminal expansion or instabilities
\cite{trodden,hsu,cline,dubovsky,buniy}, NEC violating objects such as
negative tension branes or orientifold planes can be introduced
without such problems.  In terms of energy and momentum, NEC is
violated if $\rho + p_3 <0$ or $\rho + p_d <0$ at any time or point in
space.  Equally well one could use the A-average of the above
equations to probe NEC violation \cite{SW,oxidised}.  Using the
D-dimensional Einstein equations \eref{Energy} we can then relate NEC
violation to the metric moduli and the 4d Hubble parameter
\ba
\kappa^2_{_D} \e^{2\Omega} (\rho+p_3) 
&=& -2\dot{H}
-
\sigma^2 -\frac1d \xi^2 
+2 \xi \dot \Omega + 2 \dot \Omega^2
+(H -  \partial_t)(2\dot \Omega +\xi),
\label{eq1}
\\
\kappa^2_{_D}\e^{2\Omega} (\rho+p_d)  
&=& -3 (\dot H + H^2)
- \sigma^2 
+ \frac{d+2}{d} \xi \dot \Omega + 3 H (\frac{\xi}{d} -\dot \Omega) 
-\partial_t\(3\dot\Omega +\frac{(d-1)}{d} \xi\)
\nn\\&+&\frac{(d-4)}{d} \e^{2\Omega} \nabla^2 \Omega 
+\frac{4(d-1)}{d} \e^{2\Omega} (\nabla_m \Omega) (\nabla^m \Omega)
+\frac1d  \e^{2\Omega} \R[g^{(d)}_{mn}] ,
\label{eq2}
\ea
with $\sigma^2 =\sigma_{mn} \sigma^{mn}$.  We now take the A-average
of the above equations, using that
\be
\langle 2\dot\Omega +\xi \rangle_A =0, \qquad 
{\rm with} \;\left\{ 
\begin{array}{ll}
A = 2,   \quad & ({\rm no} \; {\rm restriction}),\\
\;\forall A, & ({\rm metric} \;{\rm restriction} ),
\end{array}
\right.
\label{einsteinA}
\ee
to simplify the equations.  If no further restrictions are put on the
metric, the above equation is valid only for the canonical average $A
=2$, and follows from the definition of the Planck mass
\eref{einstein}. However, with the extra metric restriction
\eref{restrict} discussed in section \ref{s:restriction}, the equation
is valid for all values of $A$. Using this freedom allows to derive
stronger no-go theorems. We further split the fields in a zero mode
plus perturbations $Q = Q_A + Q_{A \perp}$ as in \eref{split}.  The
time derivatives of the metric function can be rewritten with
\eref{dQ}.  The Ricci scalar of the extra-dimensional manifold is
given in \eref{R}. The result is:
\ba
\kappa^2_{_D}\langle \e^{2\Omega} (\rho+p_3) \rangle_A &=& 
-2\dot H -  \langle \sigma^2 \rangle_A
-\frac{(2+d)}{2d} \langle \xi^2 \rangle_A \nn \\  
&+&\langle (2\dot \Omega +\xi)_{_{A\perp}} 
(\frac12(2\dot \Omega +\xi)_{_{A\perp}}
+(A\dot\Omega+\xi)_{_{A\perp}}) \rangle_A,
\label{result1} \\
\kappa^2_{_D}\langle \e^{2\Omega} (\rho+p_d) \rangle_A &=& 
-3 (\dot H +H^2)
- \langle \sigma^2  \rangle_A 
\nn \\ &+&
\frac{(d+2)}{2d} \[ - \langle \xi^2 \rangle_A +(3 H +\partial_t)\xi_A
+ (2 - A) \langle(\dot\Omega \xi)_{_{A\perp}}
\rangle_A  \]
\nn\\ &+&
\frac32 \langle 
(2\dot \Omega+\xi)_{_{A\perp}}(A\dot\Omega+\xi)_{_{A\perp}} \rangle_A
-C_1(A,d) \langle \e^{2\Omega} (\nabla\Omega)^2 \rangle_A +{\rm B.T.}
\label{result2}
\ea
To obtain the last term we performed a partial integration
\be
\langle \e^{2\Omega} \nabla^2 \Omega \rangle_A =
 -(2+A) \langle
\e^{2\Omega} (\nabla \Omega)^2 \rangle_A +{\rm B.T.}
\label{partial}
\ee
In the derivation of the no-go theorems it is assumed that the
boundary term 
\be
{\rm B.T.}=\langle1\rangle^{-1}\int_{\partial \M}  \sqrt{g^{(d)}}\e^{(2+A)\Omega}
\vec \Delta \Omega\cdot\dd \vec \Sigma
\label{BT}
\ee
vanishes.  Here $\vec \Sigma$ is a directed $(d-1)$-area element of
the boundary surface $\partial \M$.  The constant in \eref{result2} is
\be
C_1(A,d) = \frac1d \times\left\{
\begin{array}{ll}
A(d-4) -2d-4, \qquad\qquad & ({\rm RF}),\\
3A(d-2)-d^2+5d-10, & ({\rm CRF}),
\end{array}
\right.
\label{C1}
\ee
depending on whether the extra-dimensional space is Ricci flat (RF) or
conformal Ricci flat (CRF).  The equations are written in such a form
that it is straightforward to specialize to the two cases of interest,
namely $A=2$ which is valid for a generic metric, and $\langle
2\Omega+\xi \rangle_A = 0$ and generic $A$ for the metric with extra
restriction.

\subsection{No-go theorem}
\label{s:nogo}

To derive the no-go theorem of \cite{SW} we assume the ``metric
restriction''  (its motivation and validity is discussed in section
\eref{s:restriction}):
\be
 (2 \dot \Omega + \xi)_{_{2\perp}} =0.
\label{restrict}
\ee
Combining with the requirement of a constant Planck
mass \eref{einstein} gives $(2\dot \Omega + \xi) =0$, and thus also
\eref{einsteinA} is trivially satisfied for all $A$.  The equations
(\ref{result1},~\ref{result2}) simplify
\ba
\kappa^2_{_D}\langle \e^{2\Omega} (\rho+p_3) \rangle_A &=& 
 -2 \dot H -  |X| ,
\label{res1} \\  
\kappa^2_{_D}\langle \e^{2\Omega} (\rho+p_d) \rangle_A &=& 
-3 (\dot H + H^2)
- |X| -C_1(A,d) \langle \e^{2\Omega} (\nabla\Omega)^2 \rangle_A
-C_2(A,d) \langle \xi_{_{A\perp}}^2 \rangle_A
\nn\\&&+
\frac{(2+d)}{2d} \frac{\partial_t( a^3\xi_A)}{a^3} ,
\label{res2}
\ea
with $|X|=\langle \sigma^2 \rangle_A +\frac{(2+d)}{2d} \xi_A^2 \geq
 0$, $C_1$ given by \eref{C1}, and $C_2 = \frac{(2+d)}{4d}(4-A)$.  The
 D-dimensional matter violates the null energy condition if the
 left-hand-side of one of the above equations is negative. Depending
 on the number of extra dimensions and whether the metric is RF or
 CRF, we can choose $A$ such that the constants $C_i$ are non-negative
 and derive strong no-go theorems.  In a nutshell, 4d accelerated
 expansion \eref{dda} implies that the first term on the
 right-hand-side of \eref{res2} is negative; NEC can only be satisfied
 if this is compensated by the last term and $\partial_t \xi
 >0$. However, the growth of $\xi$ is limited, as otherwise $|X|$
 grows so large that the right-hand-side of \eref{res1} becomes
 negative and NEC is violated after all.  Note that $-\dot H  \geq 0$
 for a scale factor that grows as a power law or exponentially.

 The arguments can be made more precise. If only $C_1 \geq 0$ and
 $C_2$ arbitrary, then a four-dimensional de Sitter universe with
 $\dot H =0$ is incompatible with the NEC.  Indeed, it follows from
 \eref{res1} $\dot H =0$ is only possible for static extra dimensions
 $\xi,\sigma=0$. But this leads to a contradiction in \eref{res2} if
 $C_1 \geq 0$, which can only be resolved if NEC is violated.
 Choosing an appropriate value for $A$, $C_1$ can always made positive
 or zero except for $d=4$ respectively $d=2$ in RF and CRF. This gives
 the theorem:
\begin{itemize}
\item Theorem 1: {\it Given (1) NEC, and (2) $d \neq 4\;
   (2)$ for RF (CRF), compactification to 4d dS space is impossible.}
\end{itemize}

Stronger constraints can be derived if $C_2 \geq 0$ as well, for
example the duration of 4d accelerated expansion can be bounded.  This
requires $d < 4$ or $d \geq 10$ in RF, while for CRF only $d=2$ and $d
>14$ is excluded.  Accelerated expansion corresponds to $H^2 +\dot H
\geq 0$ \eref{dda}.  Then \eref{res1} bounds $\xi_A^2 < H^2$. The
first term of \eref{res2} is negative, which can only be positive by a
positive time-derivative
\be
a^{-3}\partial_t (a^3 \xi_A) \gtrsim \dot H + H^2
\label{growth}
\ee
In the limit that $\dot H + H^2 \approx 0$ positive but arbitrarily
small, only a slow growth of $\xi_A$ may be sufficient for the
right-hand-side of \eref{res2} to be positive, and it may take many
efolds before $\xi_A$ exceeds $H$ and NEC is violated. Hence, the
period of accelerated can only be significantly constrained if $\dot H
+ H^2 \sim H^2$ differs significantly from zero.  If the 4d theory is
general relativity to a good approximation (as is the case when
discussing dark energy, but not necessarily during inflation --- see
the discussion below \eref{FRW}), this translates to the statement
that strong bounds can be derived only when the equation of state
parameter is considerably smaller than $-1/3$ but not in the limit
$w^E \to -1/3$.  Assuming the r.h.s. of \eref{growth} is $\O(H^2)$, we
can integrate this equation to get: $\xi_A \gtrsim H^2 t + \O( H \dot
H t)$. Plugging in \eref{res1} it follows that after a time $t \sim
H^{-1}$ the NEC is violated.  We can thus formulate the following
theorem: 
\begin{itemize}
\item Theorem 2: {\it (1) Given NEC, (2) $\dot H + H^2 \sim H^2$, and (3)
  $d<4$ or $d \geq 10$ for RF and $d \neq 2$  and $d < 15$ for CRF, 4d accelerated
  expansion is only possible for a limited $\O(1)$ number
  of e-folds.}
\end{itemize}

Allowing for explicit NEC violation does not imply that accelerated
expansion is automatic.  Additional theorems can be derived that
constrain the spatial and/or temporal distribution of the NEC
violating component in the energy-momentum tensor, see
Ref.~\cite{SW} for details.

\section{Assumptions}
\label{s:assumptions}

Any no-go theorem is as good as the assumptions that go into it. Let
us therefore step back for a moment and list the assumptions that went
into the derivation of the no-go theorems in section
\ref{s:nogo}.  They are:

\begin{enumerate}[(i)]

\item 
\label{a:GRD} 
The higher dimensional theory is described by general relativity.

\item \label{a:metricD} The metric \eref{metric} is block-diagonal, with the
  higher dimensional manifold either $\R$-flat (RF) or conformal
  $\R$-flat (CRF).

\item \label{a:FRW} The 4d metric in the Einstein frame is of the FRW form with
  zero spatial curvature.

\item\label{a:restriction} The metric satisfies the restriction
  \eref{restrict}.

\item \label{a:BT} The boundary term \eref{BT} vanishes.

\item \label{a:average} The A-average is finite for all finite $A$.

\item \label{a:g_vis} The Einstein metric that brings the 4d Einstein-Hilbert action
  in canonical form is the metric of our universe.

\end{enumerate}

Condition \eref{a:GRD} states that we do not consider higher order
curvature invariants in the higher dimensional action (or consider
them negligibly small).  It also excludes possible other modifications
of gravity.  We note that dS solutions violating this assumption have
been constructed, for example, in \cite{ohta1,ohta2,ohta3,ohta4}.

It is impossible to derive no-go theorems for generic metrics.
Assumption \eref{a:metricD} simplifies the metric considerably. The
restriction to RF or CRF metrics allows to determine the sign and size
of the Ricci-scalar in \eref{eq2}.  In general, a positive/negative
extra-dimensional curvature will reduce the right-hand-side of
\eref{eq2}, making it easier/harder to satisfy the NEC.  Many models
discussed in the literature have a metric of the form
\eref{metric}. The original Kaluza-Klein (KK) model, Randall-Sundrum
(RS) models \cite{rs1,rs2} and all five-dimensional models are RF,
flux compactifications on conformal Calabi-Yau manifolds in string
theory \cite{GKP,KKLT,LVC,klebanov} are CRF.  We note that dS
solutions in set-ups with a more general extra-dimensional metric
exists, see for example\cite{neupane1,neupane2}.

Condition \eref{a:FRW} is motivated by observations. Our universe is
nearly homogeneous and isotropic on large scales, and
indistinguishable from flat \cite{WMAP}.  It excludes set-ups in which
4d Lorentz symmetry is (weakly) violated. In \cite{SW} the additional
assumption was made that the dimensionally reduced 4d theory is
described by general relativity with negligible small corrections.
This assumptions is only needed to express the theorems in terms of
the equation of state parameter $\omega^E$ rather than directly in
terms of the Hubble constant, such as we did in Theorem 2, and as
such can be lifted.

The metric is still too general to derive no-go theorems and further
assumptions, such as \eref{a:restriction} are needed. This assumption
is discussed in detail in section \ref{s:restriction} below.  As
discussed in the next subsection, the theorems only apply to
Kaluza-Klein compactifications, as brane world models generically
violate one or more of the assumptions \eref{a:BT}-\eref{a:g_vis}.
Moreover, models with negative tension branes or orientifold planes,
such as RS1 and flux compactifications, violate the higher dimensional
NEC.\footnote{If NEC violation only occurs at the boundary of
  spacetime, as in RS1, one could still try to apply the theorems to
  bulk matter only.  However, this approach fails in general because
  the boundary term is non-zero and assumption \eref{a:BT} is
  violated.}  
\subsection{Brane worlds}
\label{s:braneworlds}

To arrive at the Einstein equation \eref{result2} a partial
integration is done, where it is assumed that the boundary term
\eref{BT} vanishes, which is assumption \eref{a:BT}. This is automatic
if $\M$ is compact and closed, and thus has no boundaries (for
orbifold compactification, if the covering space has no boundaries),
as in KK compactifications. However, if $\M$ is bounded by branes,
boundary terms as the above are generically non-zero, as they are
sourced by the brane localized energy-momentum.  The partial
integration is a crucial step in deriving the no-go theorems of
section \eref{s:nogo}.  First of all, by doing so one d.o.f. is
removed as $\nabla^2 \Omega$-terms (whose sign is undetermined) are
eliminated from the equations.  Secondly, the coefficient in front of
the $(\nabla \Omega)^2$-term becomes $A$-dependent, and can be made
negative choosing a suitable $A$.  Without the partial integration,
there is no such freedom.

In co-dimension one brane worlds, for which the brane extends in $D-1$
space-time dimensions and the warp factor only depends on the
remaining coordinate, the boundary terms are given by the
Israel-junction conditions, and can be calculated explicitly. For a
higher number of co-dimensions it it is much harder to formulate the
boundary conditions, and explicit solutions to the Einstein and matter
equations only exist for co-dimension two brane worlds.  Recently the
matching conditions for co-dimension two sources in arbitrary
dimensions were formulated \cite{nierop2,nierop1}.  Specific examples
of co-dimension one and two brane worlds are 5d RS1 models and
 6d ``football shaped'' models \cite{sled1,sled2,sled3,sled4}
respectively.  In both cases, the boundary terms can be related to the
energy-momentum localized at the boundary branes, and the boundary
term appearing in the Einstein equation \eref{result2} is positive
${\rm B.T.}>0$.  Hence, the boundary term can compensate negative
terms on the right-hand-side of this equations, thereby invalidating
attempts at deriving a no-go theorem.  Although these results are for
particular set-ups, it is clear that in models where space is bounded
by branes (or more generally, localized sources of energy-momentum)
the boundary terms generically do not vanish, and its value is very
model dependent; consequently no general no-go theorems can be
derived.

As a side remark, in the older Ref. \cite{oxidised} no-go theorems
were derived under the ``boundedness'' assumption, which requires
e.g. the $A$-average $\langle \e^{2\Omega}R \rangle_A$ to be bounded
in the $A \to \pm \infty$ limit (the sign depending on the number of
dimensions).  This is in place of the RF or CRF assumption used in
\cite{SW}.  As was already noted in \cite{oxidised} boundedness is
violated by the 6d football shaped models which have the property that
the curvature diverges at the brane position.  We here note that
neglecting the boundary term \eref{BT} is another assumption that
fails for these models, and that was used in the derivation of the
theorems. This loop hole in the no-go theorems was addressed for
co-dimension two branes in \cite{nierop2}.

The extra dimension do not need to be compact and closed for the
boundary term to vanish.  It may also be that the extra dimensions are
infinite or end in a singularity, such as in RS2 and soft-wall models
respectively \cite{softwall} (where branes/domain walls are localized
inside the bulk, but not at the boundary), but that the boundary terms
vanish because of the strong warping.  However, in this case we have
to be careful that the $A$-average remains finite, and assumption
\eref{a:average} is satisfied. This is automatic for the canonical
$A=2$ average, but not for arbitrary (but finite) $A>2$.
Compactifying to an effective four-dimensional theory only makes sense
if the effective 4d Planck mass \eref{planckmass} is finite, which is
assured if the warped extra-dimensional volume --- the integral
appearing in \eref{planckmass} --- is bounded.  The Planck mass is
defined in terms of the canonical $A=2$ average \eref{average}. If we
consider arbitrary $A$-average, this only makes sense if the integrals
are finite $\langle Q \rangle_A < \infty$. This is only assured for
values $A > 2$ (or $A>0$ if any amount of warping is sufficient to
kill off the boundary terms --- but this is a model dependent
question).

Assumption \eref{a:g_vis} states that the metric of our universe $g^{\rm
  vis}_{\mu\nu} $ equals the metric $g^E_{\mu\nu}$ that brings the
dimensionally reduced 4d Einstein-Hilbert action in canonical
form. This is not automatic in brane world models.  Indeed, if the
standard model matter is localized on a brane instead of living in
the bulk the induced metric on the brane, which is the metric of our
universe, is
\be
g^{\rm vis}_{\mu\nu} = \e^{2\Omega(y_{\rm br},t)} g^E_{\mu\nu}
\ee
with the warp factor evaluated at the position of the brane $y_{\rm
  br}$. The two metrics enter because matter and gravity ``see'' a
different number of dimensions. We can normalize this to
$\Omega(y_{\rm br},0) =1$ by rescaling the the 4d coordinates. Only if
the warp factor is time-independent the two metrics coincide at all
times.  Otherwise accelerated expansion in our universe is not the
same as accelerated expansion in the canonical metric.  Writing
$a^{\rm vis} = \e^{\Omega(t)} a^E$, then $H^{\rm vis} = H^E + \dot
\Omega$ and $\dot H^{\rm vis} = \dot H^E + \ddot \Omega$ in the
Einstein equations (\ref{result1},\ref{result2}).  The extra terms can
make the r.h.s. of these equations positive, thereby circumventing the
need for NEC violation.

Finally, we would like to mention that in co-dimension one brane
worlds, it follows from the Israel-Junction conditions that if a brane
or domain wall is located somewhere in the bulk, and if it carries
energy-momentum other than a pure brane tension, 4d Lorentz symmetry
is necessarily broken \cite{george,binetruy}.  If the Lorentz symmetry
breaking is small, the model may still be phenomenologically
acceptable.  However, since assumption \eref{a:FRW} is broken, the no
go theorems do not apply.  It may be interesting to see how general
this phenomenon of 4d Lorentz symmetry breaking is in general brane
world models.

Given all the considerations of this section, it may be hard if not
impossible, to extend the no-go theorems to also include (classes of )
brane world models.

\section{The metric restriction}
\label{s:restriction}

The ``metric restriction'' \eref{a:restriction} is not listed as an
explicit assumption in Ref.~\cite{SW}, but it should be viewed as
such.  Combining \eref{restrict} with the requirement of a constant
Planck mass \eref{einstein} gives $(2\dot \Omega + \xi) =0$, and thus
\eref{einsteinA} is trivially satisfied for all $A$. The
restriction simplifies Einstein's equations, as it removes one degree
of freedom.  Moreover, it allows to write the equations unambiguously
in terms of the generalized $A$-average, which provides additional
power as a suitable $A$ can be chosen to make certain terms negative.

Ref.~\cite{SW} argues that the metric restriction is a consequence of
coordinate invariance ``in the adiabatic limit''.  It is unclear how
to interpret this statement, as $\xi$ and $\Omega$ are
time-derivatives of the metric functions, and thus also vanish in the
adiabatic limit (which makes the restriction trivially
satisfied). Since the goal is to derive no-go theorems that constrain
{\it time-dependent} metrics, to be precise metrics that give
accelerated expansion in 4d, the adiabatic limit does not seem
applicable. In the full time-dependent set-up the restriction cannot
be enforced merely by gauge invariance, as there is in general no
coordinate transformation that can set $(2 \dot \Omega +
\xi)_{_{2\perp}} =0$ \footnote{This generically requires a
  time-dependent transformation $y^m \to y^m + \delta y^m(t,y^n)$,
  which introduces an off-diagonal $\propto \dd y \dd t$ term in the
  metric.} while keeping the metric in the original block-diagonal
form \eref{metric}.

Another argument given for the metric restriction is that it removes
one linear combination of the moduli fields which in the compactified
theory has a wrong sign kinetic term.  The dimensionally reduced
Einstein-Hilbert plus matter action has kinetic terms of the form
\be 
S_{\rm kin} = \int \dd^4 x \sqrt{-g^E} \frac{1}{2 \kappa^2_4} \[ -6 H^2 +
\frac{d+2}{2d} \xi_0^2 - \frac{d-1}{d} \langle \xi_{\perp}^2 \rangle_2
-6 \langle \xi_\perp \dot \Omega_\perp \rangle_2
 -6 \langle \dot \Omega^2_{\perp}\rangle_2 \]
 + ...
\ee
where the average is the canonical $A=2$ average \eref{average}. The
ellipses denote possible matter kinetic terms. Diagonalizing the
kinetic terms, there is one linear combination of $\dot \Omega_\perp$ and
$\xi_\perp$ that has a negative kinetic term, corresponding to a ghost
mode. A constraint equation relating the fields will remove one degree
of freedom, and thus may eliminate the ghost mode. However, it is by
no means clear that the restriction \eref{restrict}, which is not
unique and is imposed by hand, is really needed to remove the ghost.
Indeed, a theory that is well defined in the extra dimensions cannot
develop a pathology merely by compactifying, that is, by integrating
over the extra dimensions. The off-diagonal components of the Einstein
equations make this happen.  They provide the necessary constraints on
the system --- they contain no second order time derivatives of the
metric functions, and thus must be imposed on the system at all times.
Only for special metrics will the restriction \eref{restrict} be a
solution of the full Einstein equations.

For the metric \eref{metric} the non-zero off-diagonal Einstein
equations are the $\{mn\}$-equation, and  $\{0m\}$:
\be
\kappa^2_{_D} T_{0m} = \nabla_p \sigma^p_m + \frac{(1-d)}{d} \nabla_m \xi
+ \frac{(d+2)}{d} \xi (\nabla_m \Omega) - 3 \partial_t \nabla_m \Omega
+2 (\nabla_p\Omega)\sigma^p_m.
\label{G0m}
\ee
The number of degrees of freedom left after gauge fixing (or
alternatively, rewriting the equations in explicitly gauge invariant
form) and applying the constraint equations depends on the symmetries
of the extra dimensional manifold. The $\{0m\}$-equation equation will
always remove one linear combination of $\Omega,\xi,\sigma$. Actually,
as soon as $\Omega$ and/or $\xi$ depend on the extra-dimensional
coordinates, they necessarily mix with some combination of $\sigma$
and possible additional matter fields, and the restriction is more
complicated than \eref{restrict}.

We can check whether the metric restriction \eref{restrict} applies to
explicit time-dependent solutions of extra dimensional models that
exist in the literature.  Einstein's equations are non-linear partial
differential equations, and explicit solutions are known only for
simple metrics. Most solutions assume a separable form of the metric,
in which all metric components can be written in the form $f(t,y) =
f_1(t) f_2(y)$.  This greatly simplifies the system, as it transforms
the Einstein equations into ordinary differential equations.  If in
addition there is only one extra dimension the equations can be solved
relatively straightforwardly.  Consider then a 5d metric, which is of
the general form \eref{metric},
\be \dd s^2 = \e^{2\Omega(t,y)} g^E_{\mu\nu}(t)\dd x^\mu \dd x^\nu 
+ b(t,y)^2 \dd y^2. 
\ee
Translating to the notation of \eref{dt_g} gives $\xi = \dot b /b$.
Ref.~\cite{gosh} tries to find solutions for $\Omega = \Omega(y)$ and
$b = b(t)$.  Since $\dot \Omega =0$ vanishes by assumption, and one
can only define the Einstein frame at all times if $\xi =0$ as well
\eref{einstein}, the metric restriction \eref{restrict} is satisfied
trivially. Also for the solutions in Ref.~\cite{kim2,kanti}, which has
$\Omega = \Omega(y)$ and $b = b(y)$, the restriction is automatic.
The solution found in \cite{binetruy} violates assumption \eref{a:FRW}
as 4d Lorentz symmetry is broken.

We have found only one explicit solution in the literature in which
the metric components have a non-separable and non-trivial form
\cite{gio}.  In this case the off-diagonal Einstein equation
\eref{G0m} and the metric restriction \eref{restrict} are
inconsistent, providing an explicit counter example to the validity of
the metric restriction.  Hence \eref{restrict} should indeed be
viewed as an extra assumption rather than a necessary condition.
Ref. \cite{gio} considers a metric of the form
\ba
\dd s^2 &=& c^2(y,\tau) (-\dd \tau^2 + \dd \vec x^2) + b^2(y,\tau) \dd y^2
\nn\\
&=& \e^{2\Omega(y,t)} (-\dd t^2 + a^2(t) \dd \vec x^2) + b^2(y,t) \dd y^2.
\label{gio}
\ea
with $a \dd \tau = \dd t$ with $\tau$ conformal time.  The split of
$c^2 = \e^{2\Omega} a(t)^2$ is chosen such that $\Omega$ satisfies
\eref{einstein}, and $a$ is the Einstein frame scale factor. In
addition the model has a bulk scalar with an appropriate potential,
such that all Einstein's equations, including the constraint equation
\eref{G0m}, are satisfied. A family of solutions is found
\be
b^2 = \eps^2 c^2, \qquad c(\tau,y) = [(\lambda(y+\tau))^{2k}+1]^{-1/(2k)},
\ee
with $\epsilon, \lambda,k$ constants that enter the potential for the
bulk field.  For this metric~\eref{gio} $\xi = \dot b/ b = \dot c/c$,
and $\dot \Omega = \dot c/c - \dot a/a$. Since $a$ depends only on
time, it contributes to the zero mode $\langle \dot \Omega \rangle_2$,
but drops out of the fluctuations. We find
\be
(\dot \Omega)_{2\perp} = \xi_{2\perp} = \(\frac{\dot c}{c}\)_{2\perp}
\qquad \Rightarrow \qquad (\dot \Omega - \xi)_{2\perp} =0.
\label{counter_ex}
\ee
And thus the restriction \eref{restrict} is not satisfied for this
solution. We checked that the Planck mass \eref{planckmass} for this
solution is finite, and thus a sensible 4d theory can be defined.
Since pathologies cannot arise merely by integrating over the 5th
dimension (unless some invalid approximations are made), the ghost
mode has been removed by the constraint equation \eref{G0m}.

\subsection{Lifting the metric restriction}
\label{s:lift}

The strong no-go theorems derived in section \ref{s:nogo} depend on
the metric restriction \eref{restrict} in two essential ways.  First
of all, the extra constraint on the metric removes one degree of
freedom. Secondly, the power of generalized $A$-averaging can be used
to choose a suitable $A$ which sets $C_i \geq 0$.  Without the metric
restriction it is not possible to derive powerful no-go theorems,
except for some simplified forms of the metric.

Lifting the metric restriction of assumption \eref{a:restriction} the Einstein
equations are only valid for $A=2$; they become:
\ba
\kappa^2_{_D}\langle \e^{2\Omega} (\rho+p_3) \rangle &=& 
-2\dot H -  |X| + \frac{(d-1)}{d} \langle \xi_\perp^2 \rangle
+6 \langle \dot \Omega_{_{\perp}}(\dot \Omega +\xi)_{_{\perp}} \rangle,
\label{resA1} \\
\kappa^2_{_D}\langle \e^{2\Omega} (\rho+p_d) \rangle &=& 
-3 (\dot H + H^2)
- |X|+ \frac{(d-1)}{d} \langle \xi_\perp^2 \rangle
+6 \langle \dot \Omega_{_{\perp}}(\dot \Omega +\xi)_{_{\perp}} \rangle
\nn\\&&
+ \frac{(d+2)}{2d} \frac{\partial_t(a^3\xi_0)}{a^3}
-C_1 \langle \e^{2\Omega} (\nabla\Omega)^2 \rangle,
\label{resA2}
\ea
where we dropped the subscript $A=2$ from the averages, and introduced
the notation $Q_0 = \langle Q \rangle$.  As before $|X|=\langle \sigma^2
\rangle + \frac{(2+d)}{2d} \xi_0^2 $. Using $A=2$ in \eref{C1} gives
$C_1 = -12/d$ for RF and $C_1 = (-d^2+11d-22)/d$ for CRF; only in the
latter case and for $3 \leq d \leq 8$ is $C_1 \geq 0$ non-negative.

No general no-go theorems can be derived. NEC requires the
left-hand-side of the above equations to be positive. Accelerated
expansion implies the first term on the right-hand-side of
\eref{resA2} is negative; but this can now always be compensated by
large $\dot \Omega_\perp$ or $\xi_\perp$ contributions, and for RF
also by the $(\nabla \Omega)^2$ term. We can only derive theorems if
we make additional assumptions on the metric, such that these
contributions are limited.

\subsubsection{Static extra dimensions}

From the work of Maldacena and Nu\~nez \cite{maldacena} we already
know that with time-independent extra dimensions the strong energy
condition has to be violated to get 4d accelerated expansion.  Can we
extend this to the weaker NEC condition?  For static extra dimensions
$|X| = \xi = \sigma_{mn} =0$.  For CRF the warp factor enters the
extra-dimensional metric and $\dot \Omega =0$ as well.  In this case
the metric restriction \eref{restrict} is trivially satisfied, and the
Einstein equation for generic $A$ (\ref{res1},~\ref{res2}) apply. If
$C_1(A,d) >0$, which only excludes $d=2$, the right-hand-side of
\eref{res2} is negative for 4d accelerated expansion and NEC is
violated. 

For RF it is harder to derive theorems as $\dot\Omega$ does not need
to vanish. The warp factor is constrained by the off-diagonal Einstein
equations, which simplify enormously for static extra dimensions. By
tuning the energy-momentum sources it may be possible to satisfy these
equations even for a time-dependent warp factor. If $T_{0m}=0$ the
warp factor needs to be time-independent $\dot \Omega =0$, the metric
restriction applies, and strong theorems can be formulated.

\begin{itemize}
\item
Theorem: {\it Given (1) NEC, (2) static extra dimensions, (3) CRF with $d \neq
  2$ or RF with $d \neq 4$ and $\dot \Omega =0$, accelerated expansion
  $\ddot a >0$ is impossible.}
\end{itemize}

\subsubsection{Generalized restriction}

Consider metrics for which the warp factor and volume modulus are
related via
\be (\dot \Omega)_\perp = f  \xi_\perp =0
\label{restriction2}
\ee
with $f$ a (time-dependent) constant. This includes the restriction
imposed by Ref.~\cite{SW} quoted in \eref{restrict} for which $f =
-1/2$, as well as the solution \eref{counter_ex} for which $f =1$
\cite{gio}.  The equations (\ref{resA1},~\ref{resA2}) become
\ba
\kappa^2_{_D}\langle \e^{2\Omega} (\rho+p_3) \rangle &=& 
-2\dot H -  |X| - C_3
 \langle {\xi_{_{\perp}}}^2 \rangle
\label{resG1} \\
\kappa^2_{_D}\langle \e^{2\Omega} (\rho+p_d) \rangle &=& 
-3(\dot H + H^2)
-|X|
+ \frac{(d+2)}{2d} \frac{\partial_t(a^3\xi_0)}{a^3}
\nn \\ &&- C_3
 \langle {\xi_{_{\perp}}}^2 \rangle
-C_1 \langle \e^{2\Omega} (\nabla\Omega)^2 \rangle
\label{resG2}
\ea
with $|X|=\langle \sigma^2 \rangle +\frac{(2+d)}{2d} \xi_0^2$. The
coefficient $C_3 = (2+d)/(2d) -(3/2)(2f+1)^2$, and $C_1$ is given as
before by \eref{C1}.  If both $C_1,C_3 \geq 0$ strong theorems
analogous to those in section \ref{s:nogo} can be derived.
This is the case for CRF with $3 \leq d \leq 8$ and
\be
\frac{-3d-\sqrt{3(d^2+2d)}}{6d} \leq f \leq \frac{-3d+\sqrt{3(d^2+2d)}}{6d}
\label{boundf}
\ee
which includes the case of the metric restriction $f = -1/2$ discussed
in section \ref{s:restriction}. The allowed interval for $f$ ranges
from $\{-0.87,-0.13\}$ for $d=3$ to $\{-0.82,-0.18\}$ for $d=8$.  For
RF $C_1 <0$ independent of the number of dimensions. We can formulate
the following theorem: 

\begin{itemize}
\item Theorem: {\it Given (1) NEC, (2) $\dot H + H^2 \sim H^2$, (3)
    extra dimensions that satisfy the generalized restriction
    \eref{restriction2} with \eref{boundf}, and (4) are CRF with $d
    \geq 3$, 4d accelerated expansion can only last for a few $\O(1)$
    efolds.}  
\end{itemize}
Of course for the special case of $f=-1/2$ the stronger theorems
derived in section \ref{s:nogo} apply.

\subsubsection{No warping}

The unwarped metric is a special case of the generalized
restriction. It corresponds to $\Omega = \Omega(t)$ and thus $\nabla_m
\Omega = \dot \Omega_\perp =0$. This gives the equations above
(\ref{resG1},~\ref{resG2}) with $f=0$ and the term proportional to
$C_1$ vanishes.  Strong theorems can be derived if $C_3 \geq 0$, which
is only the case for $d=1$ but not for higher dimensions.  The
off-diagonal Einstein equations generically also do not help to
constrain the system.  Hence

\begin{itemize}
\item Theorem: {\it Given (1) NEC, (2) $\dot H + H^2 \sim H^2$, (3) no
    warping, and (4) 5d ($d =1$), 4d accelerated expansion $\ddot a
    >0$ can only last for a few $\O(1)$ efolds.}
\end{itemize}

\section{Conclusions}

In this work we revisited no-go theorems forbidding accelerated
expansion in our four large space-time dimensions in the context of
extra-dimensional models.  The metric was taken to be of
block-diagonal form, with the extra dimensions either Ricci flat or
conformal Ricci flat. This metric allows for a non-trivial warp
factor, and applies to many of the motivated examples in the
literature.  The time-evolution of the four-dimensional Einstein
metric is constrained, as the full set-up has to satisfy the
higher-dimensional Einstein equations.

To derive the no-go theorems some rather technical and innocent
looking assumptions were made, such as the vanishing of the boundary
term \eref{BT}. Nevertheless they have far reaching consequences.
Indeed the theorems only apply to Kaluza-Klein compactifications.  As
we discussed, brane world models, or more generically set-ups with
localized sources of energy-momentum, generically violate one of the
assumptions made.

But even for KK compactifications, no-go theorems can only be derived
for simple metrics, which satisfy additional restrictions.  In
Ref.~\cite{SW} Steinhardt and Wesley (silently) assumed the a relation
between the warp factor and the volume of the extra dimensions. With
this metric restriction strong no-go theorems can be derived. For
example, if the higher dimensional NEC is satisfied 4d accelerated
expansion can only last for a couple of e-folds.

In Ref.~\cite{oxidised} the metric restriction was presented as a
necessary condition to get a well-defined four-dimensional theory. We
argue, though, that it should rather be viewed as an additional
assumption on the metric. The restriction is ad-hoc and put in by
hand.  Moreover, it may be inconsistent with the constraints coming
from the off-diagonal Einstein equations.  Indeed, the work of
Ref.~\cite{gio} provides an explicit solution to the 5d Einstein
equations that does not satisfy the metric restriction.

Any no-go theorem is as strong as the assumptions that go into it. To
obtain 4d accelerated expansion requires to violate the assumptions
made in Ref.~\cite{SW}.  The easiest way is to add branes or domain
walls to the system.  For KK compactifications, our work shows that
another way around the theorems is to look for solutions that do not
satisfy the metric restriction.

\section*{Acknowledgments}
The authors are supported by a VIDI grant from the Dutch Science
Organization NWO.  We thank Damien George for useful discussions and a
careful reading of an earlier draft.

\end{document}